\documentclass[10pt,twocolumn,amsmath,amssymb,aps,prl]{revtex4-2}

\usepackage{graphicx}
\usepackage{dcolumn}
\usepackage{bm}
\usepackage{hyperref}
\usepackage{mathtools}
\usepackage[normalem]{ulem}
\usepackage{xcolor} 

\usepackage{braket}
\usepackage{multirow}
\usepackage{comment}

\begin{document}

\title{Violation of Universal Operator Growth Hypothesis in \texorpdfstring{$\mathcal{W}_3$}{W3} Conformal Field Theories}

\author{Dileep P. Jatkar\texorpdfstring{$^{1,2}$}\ , Sujoy Mahato\texorpdfstring{$^{1}$}\ , Sukrut Mondkar\texorpdfstring{$^{1}$}\ , Praveen Thalore\texorpdfstring{$^{1,2}$}}
\email{dileep@hri.res.in, sujoymahato@hri.res.in, sukrutmondkar@hri.res.in, praveenthalore@hri.res.in}
 \affiliation{$^1$ Harish-Chandra Research Institute, Chhatnag Road, Jhunsi, Prayagraj 211019, India}
 \affiliation{$^2$ Homi Bhabha National Institute, Training School Building, BARC, Mumbai}

\date{\today}

\begin{abstract}
 We show that operator growth in large-central-charge conformal field theories with $\mathcal{W}_3$ symmetry can violate the universal operator growth hypothesis once the Liouvillian is enlarged to probe the higher-spin generators. For the generalized Liouvillian $\mathcal{L} = \kappa_1 \left( L_1 + L_{-1} \right) + \kappa_2 \left( W_2 + W_{-2} \right)$, we compute the Lanczos coefficients in the descendant module of a heavy primary and find several classes with faster-than-linear growth in the descendant level $N$, including maximally violating sectors with asymptotic behavior $b_N \sim N^2$. This superlinear growth exceeds the conjectured bound and renders the Krylov complexity divergent. We further show that the same quadratic asymptotic growth already arises in the global $SL(3, \mathbb{R})$ subalgebra, indicating that the violation is rooted in the extended higher-rank symmetry itself. Our results demonstrate that extended $\mathcal{W}$-symmetries can qualitatively modify operator growth and evade conventional bounds on information scrambling.  
\end{abstract}

\maketitle

\textit{Introduction:} Quantum complexity has emerged as a useful probe of operator growth, information spreading, and ergodicity in quantum many-body systems and quantum field theories~\cite{Baiguera:2025dkc}. In this context, Krylov complexity (K-complexity) provides a quantitative measure of how an initially simple operator spreads in the Krylov basis generated by successive action of the Liouvillian~\cite{PhysRevX.9.041017, PhysRevD.106.046007,NANDY20251}. In two-dimensional conformal field theories (CFTs), operator growth is particularly natural to study because the state-operator correspondence translates the problem into the growth of descendant states in a given module~\cite{Caputa:2021ori}. A central conjecture in this framework is the universal operator growth hypothesis (OGH), according to which the Lanczos coefficients (LC) \(b_N\) grow at most linearly with the Krylov level \(N\) in chaotic systems~\cite{PhysRevX.9.041017}. This linear growth, in turn, implies at most exponential growth of K-complexity.

{Most existing studies of the OGH have focused either on standard lattice models and other many-body systems}~\cite{Rabinovici:2021qqt, Rabinovici:2022beu, Bhattacharya:2022gbz, PhysRevE.107.024217, PhysRevE.108.054222, Bhattacharya:2023zqt, Bhattacharyya:2023dhp, PhysRevD.109.066010} or on CFT settings where the dynamics is governed by the Virasoro algebra~\cite{PhysRevD.104.L081702, Caputa:2021ori, PhysRevResearch.4.013041, PhysRevD.106.126022, Khetrapal:2022dzy, Kundu:2023hbk, Chattopadhyay:2024pdj, Caputa:2024sux,  Malvimat:2024vhr}. It is therefore natural to ask whether the same bound continues to hold in theories with extended symmetry. The physical relevance of extended algebras such as $\mathcal{W}$-algebras stretches far beyond the formal classification of two-dimensional CFTs~\cite{Bouwknegt:1992wg,Bouwknegt:1994zux}. {They are quite relevant} for a variety of strongly correlated condensed matter and statistical systems. For instance, higher-spin current algebras organize edge excitations in incompressible quantum Hall fluids~\cite{Cappelli:1992yv, Cappelli:1994wb},
 while higher-spin symmetries also emerge in SYK-like models~\cite{Peng:2018zap}. Furthermore, $\mathcal{W}$-algebras 
 play an important role in critical statistical systems,
notably manifesting as $\mathcal{W}_3$ symmetry in the critical three-state Potts model~\cite{Caldeira:2003zz} and 
in parafermionic models~\cite{PhysRevLett.70.2228}. 
Determining whether higher-spin symmetries preserve or qualitatively alter the universal growth bound is therefore an important step toward understanding operator growth beyond the Virasoro paradigm with direct relevance to condensed matter and statistical systems.

For instance, $W_\infty$ higher-spin current algebras organize the edge theory of incompressible quantum Hall fluids~[24,25]. Related non-Abelian current-algebra CFT descriptions underlie the non-Fermi-liquid fixed points of multichannel $SU(N)$ Kondo models~[26], while higher-spin symmetries also emerge in SYK-like models~[27]. Furthermore, $W$-algebras play a central role in critical statistical systems, appearing for example as $W_3$ symmetry in the critical three-state Potts model~[28] and in parafermionic models~[29].

In a CFT with only Virasoro symmetry, a natural choice of Liouvillian is \(L=\kappa(L_1+L_{-1})\), and this choice underlies the known results for Liouville theory~\cite{Caputa:2021ori}. In a theory with $\mathcal{W}_3$ symmetry, however, restricting to the Virasoro generators fails to probe the enlarged symmetry algebra. To capture the effect of the spin-3 generators on operator growth, we enlarge the Liouvillian to $\mathcal{L}=\kappa_1(L_1+L_{-1})+\kappa_2(W_2+W_{-2})$. This is the minimal choice that directly incorporates the higher-spin sector; using \(W_{\pm1}\) instead leads to the same qualitative conclusions.

In this Letter, we study operator growth in a large-central-charge CFT with $\mathcal{W}_3$ symmetry using this generalized Liouvillian. Working in the descendant module of a heavy primary, we compute the LC and show that the OGH is generically violated. In particular, we identify several classes of LC with faster-than-linear growth, including two classes whose asymptotic behavior saturates at \(b_N\sim N^2\), where $N$ is the descendent level. This superlinear growth exceeds the conjectured universal bound and implies that the corresponding K-complexity diverges. We further show that the same quadratic asymptotic growth already appears in the global $SL(3,\mathbb{R})$ subalgebra, indicating that the violation is rooted in the extended higher-rank character of the symmetry algebra itself rather than in the details of full infinite-dimensional $\mathcal{W}_3$ algebra. {Our results therefore show that extended $\mathcal{W}$-symmetry can qualitatively modify operator growth and evade known bounds on information scrambling.}

\emph{Lanczos coefficients in $\mathcal{W}_3$ CFT:} A $\mathcal{W}_3$ Toda field theory~\cite{Bouwknegt:1992wg, Bouwknegt:1994zux, Fujiwara:1998bq, Bazhanov:2001xm,Belavin:2016qaa} is a two dimensional irrational CFT that involves a spin-3 conserved current $W(z)$, and the stress tensor $T(z)$ in the presence of two background charges, $Q_1$ and $Q_2$ at infinity. We only consider the holomorphic sector; the same story repeats for the anti-holomorphic sector.  
The $\mathcal{W}_3$ Toda field theory admits a free field realization in terms of two scalar fields, $\varphi_1$, $\varphi_2$ with background charge.  The spin-2 current $T(z)$ and the spin-3 current $W(z)$ written in terms of $\varphi_1$, and $\varphi_2$ take the form
\begin{equation}
    T(z) = -\frac{1}{2}[(\partial \varphi_1)^2 \! +\! (\partial \varphi_2)^2]\! -\! \left(Q_1\, \partial^2 \varphi_1\! +\! Q_2\, \partial^2 \varphi_2\right)\ ,
\end{equation}
\begin{align}
    W(z) &= -iA \Bigg\{ \frac{1}{3}(\partial \varphi_1)^3 - \partial \varphi_1 (\partial \varphi_2)^2 \nonumber\\
    &+ \Big(Q_1 \partial \varphi_1 \, \partial^2 \varphi_1 - 2Q_2 \, \partial \varphi_1 \, \partial^2 \varphi_2 - Q_2 \, \partial \varphi_2 \, \partial^2 \varphi_2\Big) \nonumber\\
    &\quad + \left( \frac{1}{3} Q_1^2 \, \partial^3 \varphi_1 - Q_1 Q_2 \, \partial^3 \varphi_2 \right) \Bigg\}\ .
\end{align}
where, $A= 2/\sqrt{261}$.  The scalar fields $\varphi_1$ and $\varphi_2$ can be expanded in terms of their oscillator modes, which we denote as $\alpha_m^1$ and $\alpha_m^2$, respectively.  The oscillators satisfy the commutation relation
\begin{equation}
    [\alpha_m^i,\alpha_n^j] = m\delta_{ij}\delta_{m+n,0}\ .
\end{equation}
Although the Virasoro generators are at most quadratic in $\alpha_m^i$, the modes $W_n$ of $W(z)$ contain cubic terms in $\alpha^i_m$.  As a result, the commutator of Virasoro generators with the oscillator modes is linear in the oscillators, but that of the $W_3$ generators is quadratic in the oscillators.

Computation of the LC depends on the choice of the Liouvillian.  Any CFT can be studied purely in terms of the Virasoro representation theory.  This amounts to ignoring any additional symmetry of the theory and its representations.  If we do not incorporate any additional symmetry, the behavior of the LC will be identical to that observed in the Liouville CFT.  
This choice of the Liouvillian therefore cannot capture the operator growth due to new symmetry generators.
To capture this effect, we have to judiciously choose a modified Liouvillian.
The minimal such choice is $\mathcal{L} = \kappa_1 (L_{1} + L_{-1}) + \kappa_2 (W_{2} + W_{-2})$.  We could have chosen $W_{\pm1}$ instead of $W_{\pm 2}$, but there are a couple of reasons for this choice. Firstly, $W_{\pm2}$ commutes with $L_{\pm 1}$ with correlated signs, i.e., $[W_{\pm 2}, L_{\pm1} ] = 0$, while $[W_{\pm1}, L_{\pm1} ] \neq 0$ and secondly, $W_{\pm 2}$, like $L_{\pm 1}$ are the highest global generators of that symmetry.  
Nonetheless, the alternative choice of the Liouvillian, $\mathcal{L}_1 = \kappa_1 (L_{1} + L_{-1}) + \kappa_2 (W_{1} + W_{-1})$, does not alter the conclusion.  We will work
with $W_{\pm 2}$, because it provides a greater contrast compared to the Liouvillian with only Virasoro generators. 

The LC are a measure of the rate of operator growth. We start with a heavy primary state generated by a heavy operator $\Phi(z)$ which forms a representation of the $\mathcal{W}_3$ symmetry.  To evaluate operator growth, we need to look at the descendant states in the heavy operator module.  These descendant states can be created using $L_{-n}$, or $W_{-n}$, $n>0$.  The Krylov subspace is generated by $L_{-1}$ and $W_{-2}$, however, the resulting representation does not lead to orthonormal states.  In the free field realization, descendant states are expressed in terms of action of oscillator creation operators on the primary state, which is an orthonormal basis.  Thus, the oscillator realization of the states in the Krylov subspace naturally constitutes the Krylov basis. In the oscillator basis, a generic descendant state at level $N$, denoted by $\Phi_{\{m_j\}\{n_p\}}$, can be expressed in terms of the oscillator modes $\alpha_n^1$ and $\alpha_n^2$ as
\begin{widetext}
\begin{equation}
    \Phi_{\{m_j\}\{n_p\}} = \frac{1}{\mathcal{N}_{\{m_j\}\{n_p\}}} \, (\alpha_{-1}^1)^{m_1} (\alpha_{-2}^1)^{m_2} \cdots (\alpha_{-1}^2)^{n_1} (\alpha_{-2}^2)^{n_2} \cdots |\Phi\rangle,
\end{equation}
\end{widetext}
where the set of non-negative integers $ \{ \{ m_j \}, \{ n_p \} \} $ satisfying $\sum_j j m_j + \sum_p p n_p = N$ represent generalized two-partition of integer $N$.  The normalization constant is given by
\begingroup
\setlength{\abovedisplayskip}{4pt}
\setlength{\belowdisplayskip}{4pt}
\begin{equation}
    \mathcal{N}_{\{m_j\}\{n_p\}} = \left[ \prod_j m_j! \, j^{m_j} \right]^{1/2} \left[ \prod_p n_p! \, p^{n_p} \right]^{1/2}.
\end{equation}
\endgroup
To obtain LC, it is sufficient to consider the action of the operator $\kappa_1 L_{-1} + \kappa_2 W_{-2}$ in the descendant state $\Phi_{\{m_j\}\{n_p\}}$ at level $N$.  The action of remaining generators follows from the relations $L_n = L_{-n}^{\dagger}$ and $W_n = W_{-n}^{\dagger}$ among the Virasoro and spin-3 algebra generators, respectively.  The action of $W_{-2}$ on an arbitrary descendant state $\Phi_{\{m_j\} \{n_p\}}$ gives a long expression, which is provided in Eq.~\eqref{eq:lanc} of the End Matter.  This action gives rise to nineteen classes of LC, but not all of them  violate the upper bound;  among them, ten coefficients have a {faster-than-linear} growth in the large $N$ limit. For the discussion in the main text, the most relevant ones are two classes, which we will define momentarily, whose large-$N$ growth is $N^2$.

\emph{Violation of universal operator growth hypothesis:}
As can be seen from Eq.~\eqref{eq:lanc} (in the End Matter), out of the nineteen classes of LC, two classes denoted as $b^I:=b_{\{m_j\} \{n_p\} \rightarrow (\{m_{j \ne k}\},m_k-1) (\{n_{p \ne k+2}\}, n_{k+2}+1)} $ and $b^{II}:=b_{\{m_j\} \{n_p\} \rightarrow (\{m_{j \ne l+2}\}, m_{l+2}+1) (\{n_{p \ne l}\}, n_l-1)} $ {maximally violate} the conjectured upper bound.  These coefficients are obtained by acting $W_{-2}$ on a generic descendant state $\Phi_{\{m_j\}\{n_p\}}$,
    \begin{align}
        b^I &= -2iAQ_2\kappa_2\sqrt{k(k+1)^2(k+2)m_k(n_{k+2}+1)}\ ,\\
        b^{II} &= 2iAQ_2\kappa_2\sqrt{l(l+1)^2(l+2)n_l(m_{l+2}+1)}\ .
    \end{align}
In particular, we identify a specific class of descendant states whose modules lead to this violation.  At any level $N$, a descendant state denoted in terms of two sets of integers $ \{\{m_j\},\{n_p\}\}=\{ \{0, \cdots, 0, \sqrt{N}, 0, \cdots, 0 \}, \{ 0,\cdots, 0 \} \}$ or $ \{ \{ 0, \cdots, 0 \}, \{0,\cdots, 0, \sqrt{N}, 0, \cdots, 0 \} \}$, where the entry $m_j=\sqrt{N}$ appears exactly at position $j=\sqrt{N}$  or in $n_p= \sqrt{N}$ at $p=\sqrt{N}$ but not both and all other $\{m_j\}$ and $\{n_p\}$ are zero will lead to the violation of the conjectured upper bound in both $b^I$ and $b^{II}$.  
Other coefficients also violate the linear growth bound, but the dominant violation is seen in $b^I$ and $b^{II}$.  

A large class of states can be shown to maximally violate the bound.  To see this, let us consider a state at a level $N\gg 1$ in the module of the $\mathcal{W}_3$ algebra, labeled by $\{\{m_j\},\{n_p\}\}$.  In this set we will choose the subset $\{n_p\}=\{0,\cdots,0\}$ and all non-vanishing entries appear only in the subset $\{m_j\}$, and the non-vanishing entries are centered around $j=N^\alpha$ with a breadth of $N^\gamma$ in the values of $j$ and the degeneracy $m_j = N^\beta$ at each $j$ value, such that $\alpha+\beta+\gamma = 1$ and $\alpha>0,\ \beta>0,\ \gamma>0$.  In the large $N$ limit, LC $b^I$ and $b^{II}$ grow as $N^{2\alpha+(\beta/2)}$.  Whenever $\alpha-\beta/2-\gamma >0$, either the growth of $b^I$ or of $b^{II}$ is faster than $N$ and hence these coefficients violate the conjectured bound.  Since $\alpha$, $\beta$, and $\gamma$ satisfy only one constraint, the states that violate the growth bound are quite generic.  The degeneracy $\{m_j\}$ of this state reduces with every action of $W_{-2}$, but at the same time it increases the degeneracy on the $\{n_p\}$ side.  The successive action of $W_{-2}$ results in the switch of the growth pattern from $b^I$ to $b^{II}$ and back.  Therefore, either $b^I$ or $b^{II}$ grows faster than $N$ at any step.  If the action of $W_{-2}$ is continued ad infinitum, then the growth rate of LC approaches $N^2$ asymptotically.  Since the LC growth rate saturates at $N$ for CFTs with Virasoro symmetry and saturates at $N^2$ for CFTs with $\mathcal{W}_3$ symmetry, it is tempting to conjecture that for CFTs with $\mathcal{W}_n$ symmetry, the growth of LC will saturate at $N^{n-1}$.

This conclusion is further reinforced by the global subalgebra. Restricting the analysis to the global symmetry algebra $SL(3,\mathbb{R})$, which is the global part of the $W_3$ algebra, we again find quadratic asymptotic growth of the LC~\cite{SM}. Therefore, the violation of the OGH is not merely a consequence of the infinite-dimensional or nonlinear structure of the full $\mathcal{W}_3$ algebra. Rather, it already arises at the level of the extended higher-rank global symmetry itself. 


\emph{Typical descendants:} Here we address how fast the typical states grow in the regime of high descendant levels.  Correlation functions in typical high-energy states
appear thermal. In particular, we use the definition of
typical states of~\cite{Datta:2019jeo}.  
In $\mathcal{W}_3$ Toda field theory, descendants are eigenstates of number operator $\alpha^1_{-j}\alpha^1_{j} + \alpha^2_{-p}\alpha^2_p $ with eigenvalue $j m_j + p n_p$.  For a typical high-level descendant of a heavy primary, the occupation numbers $m_j$ and $n_p$ of the generalized integer two-partition $\sum_{j=1}^N j m_j + \sum_{p=1}^N p n_p = N$ are distributed according to the Boltzmann distribution and their expectation values are $\langle m_j \rangle = \frac{q^j}{1 - q^j}$, $\langle n_p \rangle = \frac{q^p}{1 - q^p}$, with $q = e^{-\pi/\sqrt{3 N}}$. 
There are three distinct large $N$ behaviors of LC for sub-modules involving typical descendants focusing on transition amplitudes with $k,l, r, q \ll N$ in Eq.~\eqref{eq:lanc}. These are $N^{1/4}$, $N^{1/2}$, and $N^{3/4}$.  The former two were also observed in \cite{Caputa:2021ori} for Liouville CFT, whereas the latter one, $N^{3/4}$, is novel.

\emph{K-complexity:} We will now investigate the effect of this growth of LC on the K-complexity. 
Since we are interested in understanding the effect of the term $W_{2} + W_{-2}$ on the K-complexity, we set $\kappa_1 =0$ in the Liouvillian.  The time evolution of a state is then given by $e^{i \kappa_2 t (W_{-2} + W_{2})} \ket{h, w}$, where $\ket{h, w}$ denotes the highest weight state with conformal weight $h$ and $\mathcal{W}_3$ charge $w$. Using the Baker–Campbell–Hausdorff (BCH) decomposition, this state can be expressed as
\begin{align}
    \ket{\mathcal{O}(t)} &= e^{i \kappa_2 t (W_{-2} + W_{2})} \ket{h, w} \nonumber \\
    &= e^{\alpha_{-} W_{-2}} e^{\alpha_{0} L_{0}} e^{\alpha_{+} W_{2}} \ket{h, w} .
\end{align}
To determine the coefficients $\alpha_{-}$, $\alpha_{+}$, and $\alpha_{0}$, one may either use the matrix representations of $W_{\pm 2}$ and $L_{0}$ \cite{Castro:2016tlm}, or perform a series expansion on both sides of the BCH decomposition and compare the coefficients of the powers of $W_{\pm 2}$ and $L_{0}$. Carrying out this computation gives
\begin{equation}
    \alpha_{+} = \alpha_{-} = \frac{i}{4} \tan(4 \kappa_2 t) ,
    \qquad
    \alpha_{0} = \log \sec(4 \kappa_2 t) .
\end{equation}
Since $\ket{h, w}$ is the highest-weight state it is annihilated by $W_{2}$, hence the time-evolved state simplifies to
\begin{equation}
    \ket{\mathcal{O}(t)} = e^{\alpha_{0} h} e^{\alpha_{-} W_{-2}} \ket{h, w} .
\end{equation}
Expanding the exponential involving $W_{-2}$ gives,
\begin{equation}
    \ket{ \mathcal{O}(t)} =
    \sum_{n=0}^{\infty}
     i^n \varphi_{n}(t)
    \ket{h + 2n,w} ,
\end{equation}
where the Krylov wavefunctions $\varphi_{n}(t)$ are given by
\begin{equation}
    \varphi_{n}(t)
    = \frac{1}{4^n \, n!}
    \sqrt{\frac{(2n)! \, \Gamma(2h+2n)}{\Gamma(2h)}}
    \frac{\tan^{n}(4 \kappa_2 t)}{\cos^{h}(4 \kappa_2 t)} .
\end{equation}
The Krylov complexity for the $\mathcal{W}_3$ CFT is then given by
\begin{align}
    K(t)
    &= \sum_{n=0}^{\infty} n \, |\varphi_n(t)|^2 \nonumber \\
    &= \sum_{n=0}^{\infty}
    \frac{n \, (2n)!}{4^{2n} \, (n!)^{2}}
    \frac{\Gamma(2h+2n)}{\Gamma(2h)}
    \frac{\tan^{2n}(4 \kappa_2 t)}{\cos^{2h}(4 \kappa_2 t)} .
\end{align}
The above series diverges, indicating that the Krylov complexity diverges for the $\mathcal{W}_3$ CFT.

\emph{Discussion and outlook:} 
Our analysis shows that the universal operator growth hypothesis, which is respected in Virasoro CFTs, can be violated once the Liouvillian is enlarged to probe higher-spin generators. The key point is not merely the infinite-dimensionality or nonlinearity of the $\mathcal{W}_3$ algebra, but the presence of an extended higher-rank symmetry sector. In particular, while Virasoro CFTs exhibit at most linear growth of Lanczos coefficients with descendant level $N$, the $\mathcal{W}_3$ case exhibits asymptotic growth up to $N^2$. Moreover, the same quadratic growth already appears in the global $SL(3,\mathbb{R})$ subalgebra. This shows that the violation is rooted in the extended higher-rank symmetry itself. These results naturally suggest that in CFTs with $\mathcal{W}_n$ symmetry, the maximal growth may scale as $N^{\,n-1}$.


At first sight, the enhancement of operator growth in a theory with more symmetry may appear counterintuitive. The resolution is that the enlarged symmetry also greatly enlarges the descendant module accessible to the dynamics. Thus, there is a competition between the additional conserved charges, which tend to constrain the dynamics, and the rapid growth of the accessible Hilbert-space sector generated by the higher-spin operators. Our results indicate that the latter effect dominates.


Gaiotto and Verlinde \cite{Gaiotto:2024kze} showed that the partition function of the DSSYK model is identical to the Schur index of the $\mathcal{N}=2$ SU(2) SYM theory~\cite{Gadde:2011uv, Gadde:2011ik}.  The AGT correspondence for 4D SU(2) theories and Liouville theory~\cite{Alday:2009aq, Alday:2009fs} was generalized, by Wyllard~\cite{Wyllard:2009hg}, to SU(N) theories and $\mathcal{W}_N$ Toda field theories.  It would be interesting to extend our results to higher rank gauge theories in 4D and generalized DSSYK models, using these correspondences.  In particular, it will be nice to identify the generalized DSSYK model, which captures the spectrum of the Schur sector of SU(3) quiver gauge theories and connects it to the $W_3$ Toda field theory studied here.


\emph{Acknowledgments :} We thank Joydeep Chakravarty, Anatoly Dymarsky, Arnab Kundu, and Giuseppe Policastro for their comments on the manuscript. We thank Pawel Caputa, Johanna Erdmenger, Suresh Govindarajan, Ren\'e Meyer, Giuseppe Policastro, and Sai Vinjanampathy for helpful discussion. DPJ acknowledges the support of the ICTP through the Associates Programme (2022-2027).

\bibliography{refs}

\appendix
\section{End Matter}

Here, we collect some relevant formulae.  
We begin with the action of $W_{-2}$ on a generic descendant state at level $N$. This expression is somewhat unwieldy and inconvenient.  For the reader's benefit, we will write it in multiple parts and later comment on each part and its relevance to the growth of LC.
\begin{equation}\label{eq:lanc}
    W_{-2}\,\Phi_{\{m_j\} \{n_p\}} = \sum_{i=1}^7 A_i \ ,
\end{equation}
where the individual parts $A_i$, $i= 1,\cdots,7$ are,
\begin{widetext}
    \begin{equation}\label{eq:Aone}
        \begin{aligned}
            A_1 =  -2iAQ_{2}&\Bigg(\sum_{k}^{(j, p-2)\not= k}\sqrt{k(k+1)^2(k+2)m_k(n_{k+2}+1)}\, \Phi_{(\{m_j\},m_k-1) (\{n_p\}, n_{k+2}+1)}\\
        &  - \sum_{l}^{(j-2,p)\not= l}\sqrt{l(l+1)^2(l+2)n_l(m_{l+2}+1)}\, \Phi_{(\{m_j\}, m_{l+2}+1) (\{n_p\}, n_l-1)}\Bigg)\ ,
        \end{aligned}
    \end{equation}
    \begin{equation}
        \begin{aligned}
            A_2 = A &\Bigg(  \sum_{k}^{j \ne (k,r,k+2-r) }\sum_{r=1}^{k+1}\sqrt{kr(k+2-r)m_{k}(m_r+1)(m_{k+2-r}+1)}\, \Phi_{(\{m_j\},m_k-1, m_{r}+1, m_{k+2-r}+1) \{n_p\}}\\ 
        &-2 \sum_{l}^{(j+q-2,p,q) \ne l}\sum_{q=1}^{l+1}\sqrt{lq(l+2-q)n_l(n_q+1)(m_{l+2-q}+1)}\, \Phi_{(\{m_j\},m_{l+2-q}+1) (\{n_p\},n_l-1,n_q+1)} \Bigg)\ ,
        \end{aligned}
    \end{equation}
    \begin{equation}
        \begin{aligned}
            A_3 = A&\Bigg( \sum_{k}^{j \ne (k,2k+2)}\sqrt{2k^2(k+1)(m_k-1)m_k(m_{2+2k}+1)}\, \Phi_{(\{m_j\},m_k-2,m_{2+2k}+1) \{n_p\}}\\
        & -  \sum_{l}^{(\frac{j-2}{2},p) \ne l}\sqrt{2l^2(l+1)(n_l-1)n_l(m_{2+2l}+1)}\, \Phi_{(\{m_j\}, m_{2+2l}+1) (\{n_p\},n_l-2)}\Bigg)\ ,
        \end{aligned}
    \end{equation}
    \begin{equation}
        \begin{aligned}
            A_4 = A&\Bigg(2 \sum_{k}^{j \ne (k,r,k+r+2)}\sum_{r>k}\sqrt{kr(k+r+2)m_km_r(m_{k+r+2}+1)}\, \Phi_{(\{m_j\},m_k-1,m_r-1,m_{k+r+2}+1) \{n_p\}}\\
        & - \sum_{k}^{\substack{(j,p+r-2) \ne k\\ p \ne r}}\sum_{r=1}^{k+1}\sqrt{kr(k+2-r)m_k(n_r+1)(n_{k+2-r}+1)}\, \Phi_{(\{m_j\},m_k-1) (\{n_p\},n_r+1,n_{k+2-r}+1)}\\
        & - 2 \sum_{k}^{\substack{(j,p-r-2) \ne k\\p \ne r}}\sum_{r}\sqrt{kr(k+r+2)m_kn_r(n_{k+r+2}+1)}\, \Phi_{(\{m_j\},m_k-1) (\{n_p\},n_r-1, n_{k+r+2}+1)}\\
        & - 2 \sum_{l}^{\substack{(j-q-2,p) \ne l\\p \ne q}}\sum_{q>l}\sqrt{lq(l+q+2)n_ln_q(m_{l+q+2}+1)}\, \Phi_{(\{m_j\}, m_{l+q+2}+1) (\{n_p\},n_l-1, n_q-1)}\Bigg)\ ,
        \end{aligned}
    \end{equation}
    \begin{equation}
        \begin{aligned}
            A_5 = A &\Bigg(\varphi_{01} \sqrt{m_{1}(m_{1}+1)} \Phi_{(\{m_{j \ne 1}\},m_1+2) \{n_p\}}
        + \varphi_{01}\sqrt{n_1(n_1+1)}\, \Phi_{\{m_j\} (\{n_{p \ne 1}\},n_1+2)}\\
        &+ 2\varphi_{02}\sqrt{(m_1+1)(n_1+1)}\, \Phi_{(\{m_{j \ne 1}\},m_1+1) (\{n_{p \ne 1}\},n_1+1)}\\ 
         & +2 \varphi_{01}\sum_{k}^{(j,j-2) \ne k}\sqrt{k(k+2)m_{k}(m_{k+2}+1)}\, \Phi_{(\{m_j\},m_{k}-1,m_{k+2}+1) \{n_p\}}\\
         &-2 \varphi_{01}\sum_{l}^{(p,p-2) \ne l}\sqrt{l(l+2)n_l(n_{l+2}+1)}\, \Phi_{\{m_j\} (\{n_p\},n_l-1,n_{l+2}+1)}\Bigg)\ ,
        \end{aligned}
    \end{equation}
    \begin{equation}
        \begin{aligned}
            A_6 = -2 A \varphi_{02}& \Bigg( \sum_{k}^{(j,p-2) \ne k} \sqrt{k(k+2)m_k(n_{k+2}+1)}\, \Phi_{(\{m_j\},m_{k}-1) (\{n_{p}\}, n_{k+2}+1)} \\
       & + \sum_{l}^{(j-2,p) \ne l}\sqrt{l(l+2)n_{l}(m_{l+2}+1)}\, \Phi_{(\{m_j\},m_{l+2}+1) (\{n_p\},n_l-1)}\Bigg)\ ,
        \end{aligned}
    \end{equation}
    \begin{equation}
                 A_7 = A\Bigg((\varphi_{01}^{2}-\varphi_{02}^{2}  + 2iQ_2\varphi_{02}) \sqrt{m_2 +1}\, \Phi_{(\{m_{j \ne 2}\},m_2+1) \{n_p\}}\!\!
     -2 \varphi_{01}(\varphi_{02}+i Q_{2} )\sqrt{n_{2}+1}\, \Phi_{\{m_j\} (\{n_{p\ne 2}\},n_2+1)}\Bigg)\ .
         \end{equation}
\end{widetext}
The notation is such that in $\Phi_{(\{m_j\},m_k-1) (\{n_p\}, n_{k+2}+1)}$, all $\{m_j\}$ and $\{n_p\}$ are left unaltered except $m_k$ and $n_{k+2}$, respectively.  
In these expressions, $\varphi_{0i}$ are zero modes of scalar fields.  The equation \eqref{eq:lanc} contains nineteen classes of LC, which are
organized in seven functions $A_i$, $i=1,\cdots, 7$ based on their growth property in the large $N$ limit.  Two classes of LC exhibit growth proportional to $N^2$ in contradiction to what is conjectured in \cite{PhysRevX.9.041017}.  These two classes of coefficients are listed in $A_1$. These two coefficients grow at the rate $N^{3/2}$ in the early steps, but asymptotically their growth saturates at $N^2$. In the main text, these two classes of LC are denoted as $b^I$ and $b^{II}$.  Two terms in $A_2$ also exhibit asymptotic growth, violating the upper bound, although the violation is not maximal in this case.  In particular, these two terms grow as $N^{3/2}$ for large $N$.  The term $A_3$ contains two terms with large $N$ growth of the LC as $N^{5/4}$.  These terms have the property that the first term is a non-terminating series, but the second one terminates.  However, when it ends, it feeds into the first term while maintaining the $N^{5/4}$ growth, in the large but finite $N$ limit.  In the strict $N\to\infty$ limit, neither of these coefficients vanishes.  The function $A_4$ contains four terms, and all LC in these terms grow as $N^{3/2}$, strictly in the limit $N\to\infty$, for generic values of $k$, $p$, $l$, and $q$.  The last three $A_i$'s, namely, $A_5$, $A_6$, and $A_7$, have coefficients that either grow linearly with $N$ or have slower growth. In particular, $A_7$ has coefficients which grow as $N^{1/2}$ in the large $N$ limit, while $A_5$ and $A_6$ have terms growing linearly with $N$.  The difference between them is that the LC in $A_5$ do not terminate, but individual coefficients in $A_6$ do with the repeated application of $W_{-2}$.  However, curiously, two terms in $A_6$ are exchanged by repeated application of $W_{-2}$, and hence their growth pattern remains unaffected.

The commutator of $\mathcal{W}_3$ with the oscillators $\alpha^i_m$ of two scalar fields $\varphi^i$, $i=1,2$ is
\begin{widetext}
    \begin{equation}\label{eq:commWn}
    \begin{aligned}
        [W_n, \alpha_m^i] &= \frac{2}{\sqrt{261}} \Bigg[ 
        -m \, \delta_{1i} \sum_p :\alpha_{n+m-p}^1 \alpha_p^1: 
        + m \, \delta_{1i} \sum_p :\alpha_{n+m-p}^2 \alpha_p^2:\\ 
        &+ 2m \, \delta_{2i} \sum_p :\alpha_{n+m-p}^1 \alpha_p^2:
         + i Q_1 m (n+2)\, \delta_{1i} \alpha_{n+m}^1 \\
        &- 2i m Q_2\{ (n+m+1) \delta_{1i} \alpha_{n+m}^2 
        + (1 - m) \delta_{2i} \alpha_{n+m}^1 \} 
        - i Q_2 (n+2) m \, \delta_{2i} \alpha_{n+m}^2 \\
        &+ n(n+1)(n+2) \left( -\frac{1}{3} Q_1 \delta_{1i} 
        + Q_1 Q_2 \delta_{2i} \right) \delta_{n+m, 0}
        \Bigg]\ .
    \end{aligned}
\end{equation} 
\end{widetext}
The Liouvillian contains two relevant operators, $L_{-1}$ and $W_{-2}$.  The action of $L_{-1}$ is already stated in the main text while discussing the Liouville theory.  In the $\mathcal{W}_3$ case, we will have two copies of the LC.  Thus, the contribution of $L_{-1}$ will give four classes of LC, but only two of them will have a maximal growth, linear in $N$. $W_{-2}$ is a more interesting operator in the Liouvillian.  The commutator \eqref{eq:commWn} is crucial in determining the action of $W_{-2}$ \eqref{eq:lanc} on a generic descendant state.

\section{Supplemental Material}

\subsection{K-Complexity and Lanczos coefficients} In the Heisenberg picture, the time evolution of the Hermitian operators is governed by the Heisenberg equation of motion.
\begin{equation}\label{eq:Heisenberg-EOM}
    \partial_t \mathcal{O}(t) = \mathrm{i} [ H, \mathcal{O}(t) ] := i \mathcal{L} \mathcal{O}(t)
\end{equation}
$\mathcal{L} \boldsymbol{\bullet} = [ H, \boldsymbol{\bullet} ]$ is the Liouvillian superoperator.  If the Hamiltonian is time-independent, the solution to the Heisenberg equation can be formally written as a series of nested commutators with the Hamiltonian.
\begin{equation}
    \mathcal{O}(t) =e^{\mathrm{i} H t} \mathcal{O}(0) e^{-\mathrm{i} H t} = e^{ \mathrm{i} \mathcal{L} t} \mathcal{O}(0)
                   = \sum_{n=0}^{\infty} \frac{(\mathrm{i} t)^n}{n !} \mathcal{L}^n \mathcal{O}(0)\ , \nonumber
\end{equation}
where, $\mathcal{L}^n \mathcal{O}(0) := [H, [H, [H, \cdots, [H, \mathcal{O}(0)] ]]$ is an $n$ times nested commutator of $\mathcal{O}(0)$ with the Hamiltonian.  
The set of all nested commutators $\{ \mathcal{L}, \mathcal{L} \mathcal{O}, \mathcal{L}^2 \mathcal{O}, \cdots \}$ span a subspace of the space of operators in the Hilbert space, known as the \emph{Krylov subspace}.  Each of the operators $\mathcal{L}^n \mathcal{O}(0)$  can be mapped to states $| \mathcal{O}_n)$ in the double Hilbert space~\cite{Magan:2020iac, Kar:2021nbm, NANDY20251} with a suitable choice of inner product, which is generally taken to be the thermal Wightman inner product \cite{NANDY20251,PhysRevD.106.046007, PhysRevX.9.041017}.  However, states $| \mathcal{O}_n)$, in general, are neither normalized nor orthogonal.  An orthonormal basis can be constructed from $\{ | \mathcal{O}_n)\}$ following \emph{Lanczos algorithm}~\cite{NANDY20251}, and the resultant orthonormal, \emph{ordered} basis $\{ | \mathcal{K}_n) \}$ is called \emph{Krylov basis}.  The Lanczos algorithm also provides a set of non-negative coefficients $\{b_n\}$, called \emph{Lanczos coefficients}, which completely determine the dynamics of the operator in the Krylov basis.  The Liouvillian superoperator $\mathcal{L}$ takes a tridiagonal form in the Krylov basis characterized by the following action,
\begin{equation}\label{eq:tridiagonal-L}
    \mathcal{L} | \mathcal{K}_n ) = b_n | \mathcal{K}_{n -1 }) + b_{n + 1} | \mathcal{K}_{n+1} )\ .
\end{equation}
The state $| \mathcal{O}(t))$ can be expanded in the Krylov basis as
\begin{equation}\label{eq:teridiagonal-L}
    |\mathcal{O}(t)) = \sum_n \mathrm{i}^n \phi_n(t)  |\mathcal{K}_{n }) \ ,
\end{equation}
with the wavefunctions $\phi_n(t)$ satisfying a discrete Schrödinger equation,
\begin{equation}
    \partial_t \phi_n(t) = b_n \phi_{n-1}(t) - b_{n+1}\phi_{n+1}(t)\ .
\end{equation}
The localized initial state $| \mathcal{O}(0))$ grows in size in the Krylov basis during time evolution, and the rate of this growth is quantified in terms of the LC.  A simple localized initial state becomes more and more ``complex'' during time evolution due to the growth of its support.  This notion of operator or state growth is captured by the Krylov complexity (K-complexity),
\begin{equation}
    K_{\mathcal{O}}(t) := \sum_n n | \phi_n(t) |^2\ .
\end{equation}
A universal operator growth hypothesis (OGH) put forth by \cite{PhysRevX.9.041017} conjectured that the fastest operator or state growth corresponds to linear growth of Lanczos coefficients $b_n \sim n$, which translates into exponential growth of K-complexity.


\subsection{Lanczos coefficients in Liouville CFT}

The Liouville theory is a typical example of a large central charge irrational CFT with Virasoro symmetry.  LC and K-complexity have been computed for heavy primaries in the Liouville CFT defined on a cylinder~\cite{Caputa:2021ori}.  The Liouville theory is a theory of a single scalar field with a background charge.  While the Krylov basis is a subset of the basis provided by the oscillator modes of the free scalar field, the Liouvillian is given in terms of the Virasoro generators, $\mathcal{L} = \kappa ( L_{1} + L_{-1})$, where $L_{\pm 1}$ are the global Virasoro generators.  The mode expansion of the Liouville field $\varphi(z)$ can be stated in terms of the current,
\begin{equation*}
    \partial \varphi(z) = -i \sum_{m = -\infty}^{\infty} \alpha_m z^{-m-1}\ .
\end{equation*}
The operators $\alpha_m$ satisfy the usual oscillator commutation relation, and the stress tensor of the Liouville CFT is given by 
$T(z) = -(1/2)\partial \varphi \partial \varphi + Q \partial^2 \varphi$,
where \( Q \in \mathbb{R} \) is the background charge.  The modes of the stress tensor satisfy the Virasoro algebra with central charge $c = 1 + 12Q^2$.  
A descendant state at level $N$, $\Phi_{\{m_i\}}$, can be expressed in terms of oscillators $\alpha_n$ as
\begin{equation}
\begin{aligned}
    \Phi_{\{m_i\}} &= \frac{1}{\mathcal{N}_{\{m_i\}}} \alpha_{-1}^{m_1} \alpha_{-2}^{m_2} \cdots |\Phi\rangle,  \\
    \mathcal{N}_{\{m_i\}} &= \sqrt{\prod_j m_j! \, j^{m_j}}\ ,
    \end{aligned}
\end{equation}
where ${m_j}$ satisfying $\sum j m_j = N$ are partitions of integer $N$, denoted as $p(N)$ and $|\Phi\rangle$ is a state created by a heavy primary $\Phi(z)$.  The Lanczos coefficients were originally defined in terms of the moments of the autocorrelation function~\cite{PhysRevX.9.041017}.  However, computation of the autocorrelation function is not always feasible, and it also misses details of operator growth in systems with extended symmetries.  Therefore, following~\cite{PhysRevResearch.4.013041}, we will use the generalized definition of the LC using the action of raising and lowering operators of the CFT on the Krylov basis.

In the oscillator basis, the action of $\mathcal{L}$ on an arbitrary descendant at level $N$ generates a pair of new descendants, one at level $N+1$ and another at level $N-1$.  Due to the degeneracies of the Verma module, the LC generalize to the Lanczos matrices $b_{ \{m_i\} \rightarrow \{ r_j \} }$.  A Lanczos matrix is a rectangular matrix of dimension $p(N) \times p(N+1)$. Lanczos matrix elements~\footnote{The terms \emph{Lanczos matrix elements} and \emph{Lanczos coefficients} are used interchangeably in this manuscript.} can be obtained by analyzing the action of $L_{-1}$ on an arbitrary descendant.  The Lanczos matrices associated with the action of $L_{1}$ can be derived using the property $L_1 = L_{-1}^{\dagger}$ of the Virasoro generators. 

Applying the Virasoro generator \( L_{-1} \) to \( \Phi_{\{m_i\}} \), we get two classes of Lanczos coefficients~\cite{Caputa:2021ori}:
\begin{align}
 b_{\{m_i\} \rightarrow \{\ldots, m_n - 1, m_{n+1} + 1, \ldots\}}^{(1)} 
&= \kappa \sqrt{n(n+1)m_n (m_{n+1} + 1)} \nonumber\\
 b^{(2)}_{\{m_i\} \rightarrow \{m_1 + 1, m_2, \ldots\}} 
&= \kappa \varphi_0 \sqrt{m_1 + 1}.\nonumber
\end{align}
In the large $N$ limit, while maximal growth of $b^{(1)}$ is linear in $N$, that of $b^{(2)}$ is at most $N^{1/2}$.

\section{Lanczos Coefficient for \texorpdfstring{$SL(3,\mathbb{R})$}{SL(3,R)}}
To support our previous results on asymptotic growth of the Lanczos coefficient for the $\mathcal{W}_3$ CFT, we calculate the LC in the $SL(3,\mathbb{R})$ case, which constitutes the global subalgebra of $\mathcal{W}_3$. We work with the highest-weight representation of the $SL(3,\mathbb{R})$ group, adopting the matrix representation and conventions detailed in~\cite{Castro:2016tlm}. Before proceeding further, we review the principal embedding of $SL(N,\mathbb{R})$.

\subsection{Principal Embedding of \texorpdfstring{$SL(N,\mathbb{R})$}{SL(N,R)}}

A convenient basis for the $SL(N,\mathbb{R})$ algebra is represented by {$L_0, L_{\pm 1}$}, the generators of $SL(2,\mathbb{R})$ subalgebra, and the generators of the higher spin generators, $W_j^{(s)}$, with $j = -(s-1), \cdots, (s-1)$. Their commutation relations are as follows 

\begin{align}
    [L_{i}, L_{i^{\prime}}] & = \, (i - i^{\prime}) \, L_{i + i^{\prime}},\\
        [L_{i},W_{j}^{(s)}] \, & = \, (i(s-1) - j) \, W_{i+j}^{(s)}.
\end{align}

In this notation, $L_0$ and $W_0^{(s)}$ are elements of the Cartan subalgebra, the rest of the generators are raising and lowering operators. These commutation relations represent the principal embedding of $SL(N,\mathbb{R})$. We can explicitly represent the other $SL(N,\mathbb{R})$ generators in terms of the $SL(2,\mathbb{R})$ generators, regardless of the representation, as follows:

\begin{align}
    W_{j}^{(s)} &= \, (-1)^{s-j-1} \frac{(s+j-1)!}{(2s-2)!} \nonumber\\
    &\times \underbrace{[L_{-1},[L_{-1}, \ldots, [L_{-1}}_{s-j-1 \, \text{terms}} \, ,L_{1}^{s-1}] \ldots ]].
\end{align}

We write the highest-weight state of $SL(N,\mathbb{R})$ algebra as $\ket{hw} \equiv \ket{h,w_3, \ldots, w_N}$ with the following properties:

\begin{equation}
    \begin{aligned}
        L_{0}\ket{hw} &= h\,\ket{hw}, & L_{1}\ket{hw} &= 0,\\
        W^{(s)}_{0}\ket{hw} &= w_s\,\ket{hw}, & W^{(s)}_{j}\ket{hw} &= 0,\ j=1,\ldots,s-1 .
    \end{aligned}
\end{equation}

Where the constant $h$ and $w_s$ with $s = 3, \ldots, N$ are the parameters defining the representation. A descendant state is created by acting the raising operators: $L_{-1}$ and $W_{-j}^{(s)}$.

\subsection{N = 3}
For $SL(3,\mathbb{R})$, we have $8$ generators labeled as $T_a = \{L_i , W_m \}$ with $i = -1,0,1$ and $m = -2 , \ldots , 2$. The algebra then becomes 

\begin{equation}
    \begin{aligned}\label{w3_algebra}
        [L_i, L_j] &= (i - j)L_{i+j}\\
        [L_i, W_m] &= (2i - m)W_{i+m}\\
        [W_m, W_n] &= -\frac{1}{3}(m-n)(2m^2 + 2n^2 - mn 8)L_{m+n}
    \end{aligned}
\end{equation}

We can express the higher spin generators ${\mathcal{W}_m}$ in terms of $\mathfrak{sl}(2,\mathbb{R})$ generators as follows:

\begin{equation}
    W_m = (-1)^{2-m} \frac{(2+m)!}{4!}\underbrace{[L_{-1},[L_{-1}, \ldots, [L_{-1}}_{(2-m)\, \text{terms}} \, ,L_{1}^{2}] \ldots ]]
\end{equation}

Explicitly, we can write
\begin{align}
    W_2 &= L_1^2,\\
    W_1 & = \frac{1}{2}(L_{1}L_{0} + L_{0}L_{1}),\\
    W_0 & = \frac{1}{6}(L_{1}L_{-1} + L_{-1}L_{1} + 4 L_{0}^2),\\
    W_{-1} & = \frac{1}{2}(L_{-1}L_{0} + L_{0}L_{-1}),\\
    W_{-2} & = L_{-1}^{2}.
\end{align}

We write the highest-weight state as $\ket{h,w}$ with the following properties: 

\begin{equation}
    \begin{aligned}
        L_{0}\ket{h,w} &= h\,\ket{h,w}, & L_{1}\ket{h,w} &= 0,\\
        W_{0}\ket{h,w} &= w\,\ket{h,w}, & W_{j}\ket{h,w} &= 0,\qquad j=1,2 .
    \end{aligned}
\end{equation}

The action of $L_{-1}$ and $W_{-1}$ on the highest-weight state give descendant states at $level-1$, while the action of the $W_{-2}$ gives a descendant at $level-2$ i.e. 

\begin{equation}
    \begin{aligned}
        L_{-1}\ket{h,w} & \sim \ket{h+1,w},\\
        W_{-1}\ket{h,w} & \sim \ket{h+1,w},\\
        W_{-2}\ket{h,w} & \sim \ket{h+2,w}.
    \end{aligned}
\end{equation}

Consider a descendant state at $level-n$, which is given by $\ket{h+n,w}$. We act $L_{-1}$, $W_{-1}$, and $W_{-2}$ on this state to obtain the LC. Their action is as follows:

\begin{equation}
    L_{-1}\ket{h + n,w} = \sqrt{(n + 1)(2h + n)} \, \ket{h + (n+1), w},
\end{equation}

\begin{align}
    W_{-1} \ket{h + n, w} &= \frac{1}{2}(L_{-1}L_{0} + L_{0}L_{-1}) \ket{h + n, w}\nonumber\\
    &\hspace{-1cm} = (h + n + \frac{1}{2})\sqrt{(n + 1)(2h + n)} \nonumber\\
    &\times \, \ket{h + (n+1), w}, 
\end{align}

\begin{align}
    W_{-2} \ket{h + n, w} &= L_{-1}^2 \ket{h + n, w}\nonumber\\
    &\hspace{-1.5cm}= \sqrt{(n+1)(n+2)(2h+n)(2h+n+1)} \nonumber\\
    &\times \, \ket{h + (n+2), w}.
\end{align}

Therefore, the Lanczos coefficients are 

\begin{align}
    b_n^1 &= \sqrt{(n + 1)(2h + n)},\\
    b_n^2 &= (h + n + \frac{1}{2})\sqrt{(n + 1)(2h + n)},\\
    b_n^3 &= \sqrt{(n+1)(n+2)(2h+n)(2h+n+1)}.
\end{align}

In the large $n$ limit, $b_n^1$ shows linear growth in $n$ while $b_n^2$ and $b_n^3$ grow quadratically in $n$. When we consider the full $\mathcal{W}_3$ group, there are more types of Lanczos coefficients than these but the highest growth rate is still $n^2$.


\end{document}